\documentclass[aip,graphicx]{revtex4-1}
\usepackage{amssymb,amsmath,amsfonts,latexsym}
\usepackage{graphicx,color}
\usepackage[english]{babel}
\usepackage{tabularx}
\usepackage{mathtools}
\usepackage{mathrsfs} 
\usepackage{graphicx, texdraw}
\newcommand{\vect}[1]{\boldsymbol{#1}}

\begin{document}

\title{Tunable critical Casimir forces counteract Casimir--Lifshitz attraction}

\author{Falko Schmidt}
\email{schmidt.falko@uni-jena.de}
\affiliation{Department of Physics, University of Gothenburg, SE--41296 Gothenburg, Sweden}
\affiliation{Institute of Applied Physics, Friedrich-Schiller-University, D--07745 Jena, Germany}

\author{Agnese Callegari}
\affiliation{Department of Physics, University of Gothenburg, SE--41296 Gothenburg, Sweden}

\author{Abdallah  Daddi-Moussa-Ider}
\affiliation{Institut f\"{u}r Theoretische Physik II: Weiche Materie, Heinrich-Heine-Universit\"{a}t D\"{u}sseldorf, 
D--40225 D\"{u}sseldorf, Germany}
\affiliation{Abteilung Physik lebender Materie, Max-Planck-Institut f\"{u}r Dynamik und Selbstorganisation, 
D--37077 G\"{o}ttingen, Germany}

\author{Battulga Munkhbat}
\affiliation{Department of Physics, Chalmers University of Technology, SE--412 96 Gothenburg, Sweden}
\affiliation{Department of Photonics Engineering, Technical University of Denmark, DK--2800, Kgs. Lyngby, Denmark}

\author{Ruggero Verre}
\affiliation{Department of Physics, Chalmers University of Technology, SE--412 96 Gothenburg, Sweden}

\author{Timur Shegai}
\affiliation{Department of Physics, Chalmers University of Technology, SE--412 96 Gothenburg, Sweden}

\author{Mikael K\"{a}ll}
\affiliation{Department of Physics, Chalmers University of Technology, SE--412 96 Gothenburg, Sweden}

\author{Hartmut L\"{o}wen}
\affiliation{Institut f\"{u}r Theoretische Physik II: Weiche Materie, Heinrich-Heine-Universit\"{a}t D\"{u}sseldorf, 
D--40225 D\"{u}sseldorf, Germany}

\author{Andrea Gambassi}
\affiliation{SISSA -- International School for Advanced Studies, IT--34136 Trieste, Italy}
\affiliation{INFN, Sezione di Trieste, IT--34136 Trieste, Italy}

\author{Giovanni Volpe}
\email{giovanni.volpe@physics.gu.se}
\affiliation{Department of Physics, University of Gothenburg, SE--41296 Gothenburg, Sweden}

\date{\today}


\begin{abstract}
Casimir forces in quantum electrodynamics emerge between microscopic metallic objects because of the confinement of the vacuum electromagnetic fluctuations occurring even at zero temperature. 
Their generalization at finite temperature and in material media are referred to as Casimir--Lifshitz forces.
These forces are typically attractive, leading to the widespread problem of stiction between the metallic parts of micro- and nanodevices.
Recently, repulsive Casimir forces have been experimentally realized but their reliance on specialized materials prevents their dynamic control and thus limits their further applicability.
Here, we experimentally demonstrate that repulsive critical Casimir forces, which emerge in a critical binary liquid mixture upon approaching the critical temperature, can be used to actively control microscopic and nanoscopic objects with nanometer precision. 
We demonstrate this by using critical Casimir forces to prevent the stiction caused by the Casimir--Lifshitz forces.
We study a microscopic gold flake above a flat gold-coated substrate immersed in a critical mixture. Far from the critical temperature, stiction occurs because of dominant Casimir--Lifshitz forces. Upon approaching the critical temperature, however, we observe the emergence of repulsive critical Casimir forces that are sufficiently strong to counteract stiction. 
This experimental demonstration can accelerate the development of micro- and nanodevices by preventing stiction as well as providing active control and precise tunability of the forces acting between their constituent parts.
\end{abstract}

\maketitle

\section{Introduction}

Long-ranged forces emerge on microscopic objects when a fluctuating field is spatially confined by them, irrespective of their actual nature. Quantum-electro-dynamical (QED) Casimir forces, for example, act on neighboring uncharged conducting objects because they effectively confine electromagnetic fluctuations of the quantum vacuum\cite{Casimir1948}. Their generalization at finite temperature and in material media are known as Casimir--Lifshitz forces \cite{london1937general,Casimir1948,lifshitz1956theory}. Since these forces are attractive, they can cause the well-known problem of stiction between the metallic parts of nanodevices, such as those found in micro- and nanoelectromechanical systems (MEMS and NEMS) \cite{maluf2004}. This has motivated several recent studies to search for repulsive Casimir--Lifshitz forces both in vacuum \cite{leonhardt2007,milton2012,camacho2021engineering} and in media \cite{inui2011,pappakrishnan2014}.
However, these repulsive forces have only been obtained for specifically-engineered systems, e.g., by a careful choice of the materials of the two interacting surfaces and the separating liquid \cite{munday2009measured}, by using metamaterials with negative refractive index that are not readily available \cite{leonhardt2007}, by coating one of the surfaces with a low-refractive-index material \cite{zhao2019}, or by adjusting the concentration of ligands in solution \cite{munkhbat2021tunable}. Recently, experimental studies have shown that a combination of attractive and repulsive Casimir--Lifshitz forces can be used to levitate particles away from a surface \cite{zhao2019}, and that they can trap freely floating particles near surfaces leading to the formation of Fabry-P\'{e}rot cavities when multiple particles assemble on-top of each other \cite{munkhbat2021tunable}. Despite these advances in the manipulation of nanoparticles, these systems are inherently static and lack any tunable system parameters to allow dynamic control.

Critical Casimir forces are the thermodynamic analogue of QED Casimir forces and were first theoretically predicted by Fisher and de Gennes in 1978 \cite{fisher1978}. They arise, for example, between objects immersed in a binary liquid mixture kept near a second-order phase transition, due to the confinement of the thermal fluctuations of the local concentration of one of the components of a binary mixture (which is the order parameter of the transition) \cite{gambassi2009casimir}.
In soft matter, thermal fluctuations typically occur on molecular length scales (subnanometer) and, accordingly, the effects they produce are generally negligible at larger separations; however, upon approaching the critical point of a second-order phase transition, the fluctuations become correlated on much longer length-scales (up to several microns) and thus can significantly affect the behavior of microscopic systems \cite{krech1994,gambassi2009casimir}.
Importantly, critical Casimir forces can be either attractive or repulsive depending on the surface chemistry of the involved objects \cite{fukuto2005critical, Hertlein2008, gambassi2009critical}: 
While the critical Casimir forces between surfaces with similar adsorption preferences (either hydrophilic or hydrophobic) are attractive, they become repulsive when these preferences are opposite.

The first direct experimental measurement of critical Casimir forces was achieved only in 2008 \cite{Hertlein2008}: Using total internal reflection microscopy (TIRM), femtonewton forces were measured on a single spherical microscopic particle facing a planar surface, both being immersed in a critical water--lutidine mixture. Since then, these forces have been investigated under various conditions, e.g., by varying the properties of the involved surfaces, demonstrating the occurrence of both attractive and repulsive forces \cite{gambassi2011} as well as the existence of many-body effects \cite{paladugu2016}. Studies of the phase behavior of colloidal dispersions in a critical mixture \cite{gambassi2009critical,maciolek2018collective}
have indicated critical Casimir forces as viable candidates to control the self-assembly of micro- and nanostructures \cite{faber2013controlling,Nguyen2016,nguyen2017switching} and quantum dots \cite{marino2016}.
In addition, fluctuation-induced effects similar to critical Casimir forces have been investigated, e.g., at the percolation transition of a chemical sol \cite{gnan2014casimirlike}, in the presence of temperature gradients \cite{najafi2004forces}, in granular fluids \cite{cattuto2006fluctuationinduced}, and in active matter systems \cite{ray2014casimir,lee2017fluctuation}.

Here, we experimentally demonstrate that repulsive critical Casimir forces can be used to compensate for attractive Casimir--Lifshitz forces.
In particular, we study a microscopic gold flake diffusing above a gold surface immersed in a critical binary liquid mixture.
We control the attractive/repulsive character of the critical Casimir forces by chemical functionalization of the gold flake and surface, obtaining repulsive forces for opposite surface functionalizations, and the magnitude of these forces by varying the temperature of the mixture.
Far from the critical temperature, we observe stiction between the flake and the surface due to Casimir--Lifshitz forces.
However, as we approach the critical temperature, we observe the emergence of repulsive critical Casimir forces that are sufficiently strong to prevent stiction and to release the flake from the surface, above which it then levitates.
In addition, we demonstrate that this behavior is reversible and can be employed to actively control the flake position above a structured surface.

\section{Results}

\subsection{Working principle}

\begin{figure*}[b!]
    \includegraphics[width=\textwidth]{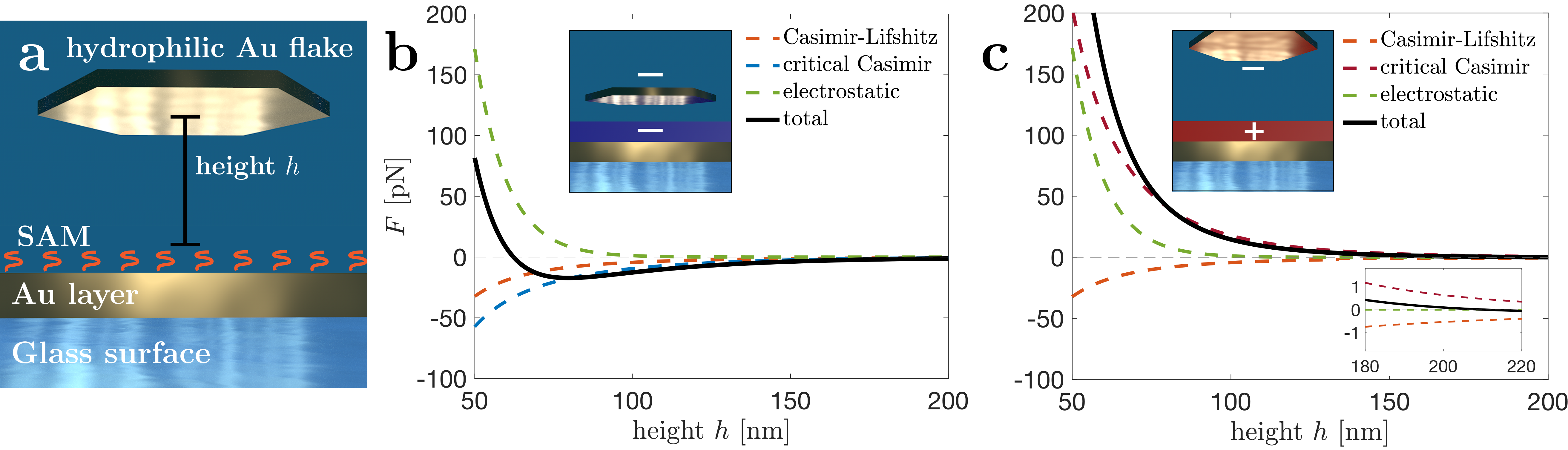}
    \caption{
    {\bf Casimir--Lifshitz and critical Casimir forces between parallel plates.} 
    {\bf a.} Schematic (not to scale) of a hydrophilic ($-$) gold flake hovering at an equilibrium height $h$ above a glass surface coated with a gold layer and treated with self-assembled monolayers (SAM) to control the preferential surface adsorption. 
    {\bf b.} Forces acting on a hydrophilic flake above a hydrophilic (blue layer, inset) surface as a function of its height $h$. Since the boundary conditions are symmetric $(-, -)$, both the Casimir--Lifshitz forces (dashed orange line, Eq.~\eqref{eq:freenergy_2sl_short_goldcoated}) and the critical Casimir forces (dashed blue line, Eq.~\eqref{eqn:CCF} at $\Delta T = T-T_{\textrm{c}}=-0.1\,{\rm K}$, with $T$ the solution temperature and $T_{\textrm{c}}$ the critical temperature for the critical binary mixture) are attractive. The total force (black line) including also a repulsive electrostatic component (dashed green line) vanishes at $h\approx80\,{\rm nm}$.
    {\bf c.} Forces on a hydrophilic $(-)$ flake above a hydrophobic $(+)$ surface (red layer, top inset) as a function of $h$. Here, the antisymmetric $(-, +)$ boundary conditions induce a repulsive critical Casimir force (dashed red line,  Eq.~\eqref{eqn:CCF} at $\Delta T = -0.1\,{\rm K}$), while the Casimir--Lifshitz forces (dashed orange line, Eq.~\eqref{eq:freenergy_2sl_short_goldcoated}) remains attractive. Accordingly, the total force vanishes at much larger $h\approx210\,{\rm nm}$ (see bottom inset). The presence of repulsive critical Casimir forces raises the equilibrium height of the flake above the surface greatly.
    The forces shown in {\bf b} and {\bf c} are calcualted for a 34-nm-thick 1520-nm-wide gold flake suspended in water--2,6-lutidine above a 40-nm-thick gold layer.
    }
    \label{fig1}
\end{figure*}

We first describe a simple model to clarify how repulsive critical Casimir forces can be employed to compensate for attractive Casimir--Lifshitz forces, by estimating theoretically the orders of magnitude of the involved forces.
We consider two flat parallel plates, as shown in Fig.~\ref{fig1}a: One is a thin gold flake (length $a=1520\,{\rm nm}$, thickness $b=34\,{\rm nm}$) suspended in a critical binary mixture above a gold-coated (thickness 40\,nm) glass substrate, which is the second plate.
The first plate hovers at a height $h$ above the surface because of the interplay of three forces (see details in Methods~\ref{sec:forces}): 
the Casimir--Lifshitz force, which, in this case, is attractive; the critical Casimir force, the character and strength of which can be varied by modifying on the surface functionalization and tuning the solvent temperature; and the electrostatic force, which provides a short-range repulsion between the  flake and the substrate. The gravitational force acting on the flake is negligible compared to the other forces acting on the flake in close proximity to the substrate.

The attractive Casimir--Lifshitz force between two metallic plates depends on $h$ as
\begin{equation}\label{eqn:CF}
    F_{\rm CL}(h) = -\frac{{\rm d}G_{\rm CL}}{{\rm d}h} \cdot S,
\end{equation}
where $G_{\rm CL}$ is the Casimir--Lifshitz free energy per unit area of the system and $S$ is the area of the flake (Ref.~\citenum{parsegian2006} and  Methods~\ref{sec:forces}). 
In this case, the force depends on the thicknesses of the flake and of the gold layer deposited on the glass surface, as well as on the dielectric properties of the glass and of the solvent (Methods~\ref{sec:forces} and Supplementary~Fig.~S1). 
The resulting Casimir--Lifshitz forces are plotted as dashed orange lines in Figs.~\ref{fig1}b,$\,$c for two opposite surface treatments of the gold layer.

The critical Casimir forces are induced by the confinement of the critical mixture between the flake and the substrate. 
Their strength $F_{\rm crit}$ depends on the adsorption preference of the involved surfaces and on the difference $\Delta T = T-T_{\rm c}$ between the actual temperature $T$ of the binary mixture and its critical temperature $T_{\rm c}$. In particular, 
\begin{equation}\label{eqn:CCF}
	F_{\rm crit}(h,\Delta T)
	= 
	-\frac{k_{\rm B}T_{\rm c}}{h^{3}}\, \theta\left({h}/{\xi(\Delta T)} \right) \cdot S,
\end{equation}
where $\xi(\Delta T)$ is the correlation length of the order parameter of the fluctuations of the mixture depending on $\Delta T$, and $\theta\left({h}/{\xi(\Delta T)} \right)$ is a universal scaling function \cite{gambassi2009casimir}. For the sake of simplicity, in the following we will omit the explicit dependence on $\Delta T$ of $\xi$.
Whether the critical Casimir forces are repulsive or attractive depends on the boundary conditions given by the surface preferential adsorption of one of the two components of the mixture (indicated by $-$ or $+$).
We can control these boundary conditions by chemically functionalizing the surfaces with appropriate self-assembled monolayers (SAMs, see Methods~\ref{sec:ExpDet}). 
In the case of symmetric boundary conditions $(-,-)$ or $(+,+)$ (i.e., the flake and surface prefer the same component of the mixture), the critical Casimir forces are attractive (dashed blue line in Fig.~\ref{fig1}b). 
For antisymmetric boundary conditions $(-,+)$ (i.e., preference for different components), the critical Casimir forces are repulsive (dashed red line in Fig.~\ref{fig1}c).

The electrostatic repulsive force $F_{\rm ES}$ is due to the double layer of counterions in the near proximity of the surface of the flake and of the substrate\cite{israelachvili2011intermolecular}. According to DLVO theory \cite{derjaguin1987DLVO} (neglecting boundary effects), we have
\begin{equation}\label{eqn:ESforce}
	F_{\rm ES}(h)
	= P_0\,e^{-h/\lambda_{\rm D}}
	\cdot S,
\end{equation}
where $\lambda_{\rm D}$ is the Debye screening length of the critical mixture, and $P_0$ is a parameter with the dimension of a pressure. 

The theoretical predictions presented in Figs.~\ref{fig1}b,c show that, in the configuration we consider at $\Delta T = -0.1\,{\rm K}$, the magnitude of the critical Casimir forces is expected to be larger than that of the Casimir--Lifshitz forces so that we can use the critical Casimir forces to tune the equilibrium position of the flake above the surface.
In the case of asymmetric boundary conditions $(-,+)$ (Fig.~\ref{fig1}c), this provides an additional repulsion between flake and surface, which can be used to prevent stiction.

\subsection{Experimental setup and analysis}

\begin{figure*}
    \includegraphics[width=.8\textwidth]{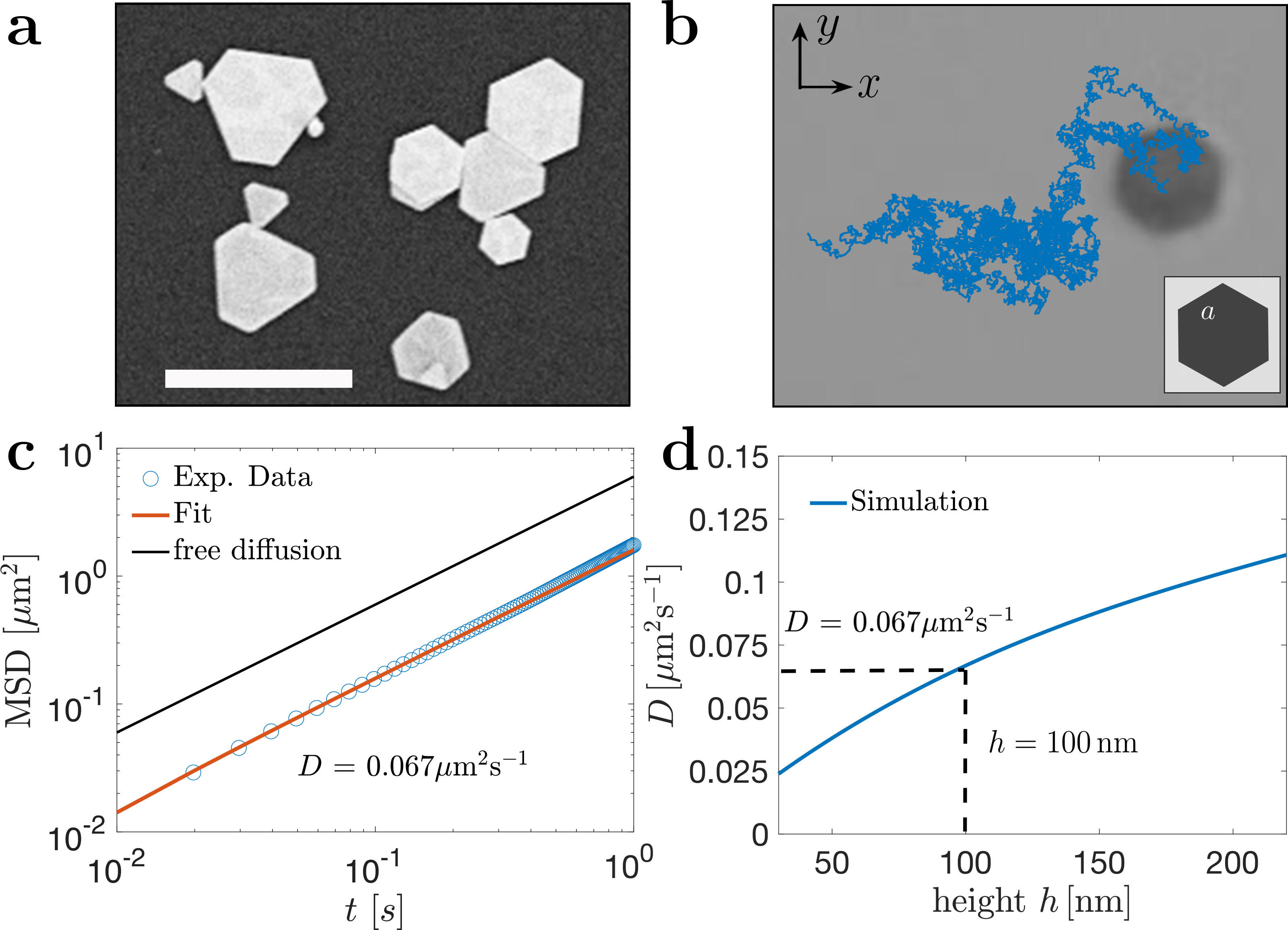}
    \caption{
    {\bf Hydrodynamic measurement of flake height above the substrate.} 
    {\bf a} Scanning electron microscope image of gold flakes with various sizes and shapes. The scale bar represents $5\,{\rm \mu m}$.
    {\bf b} Microscope image and trajectory recorded for 100\,s of a diffusing flake shaped as equilateral hexagon with side $a = 840\,{\rm nm}$ (inset) in a critical water--2,6-lutidine mixture at $\Delta T=-0.5\,{K}$ (Supplementary Video 1). 
    {\bf c} Experimental data (blue circles) and theoretical fitting (orange line) of the mean square displacement (MSD) of the flake's trajectory in \textbf{b}, which provides an estimate of the diffusion constant $D= 0.067\,{\rm \mu m^2 s^{-1}}$ compared to free diffusion with D $\sim 5.98\,{\rm \mu m^2 s^{-1}}$ (black line).
    {\bf d} Theoretical diffusion $D$  of a hexagonal flake with side $a = 840\,{\rm nm}$ as a function of its height $h$ above the surface (see Fig.~\ref{fig1}a) obtained from hydrodynamic simulations. The experimentally measured $D= 0.067\,{\rm \mu m^2 s^{-1}}$ corresponds to height $h = 100\,{\rm nm}$.
    }
    \label{fig2}
\end{figure*}

In the experiment we consider a gold flake suspended in a near-critical water--2,6-lutidine mixture (lutidine critical concentration $c_{\rm L}^{\rm c}=0.286$ mass fraction, lower critical temperature $T_{\rm c} \approx 310\,{\rm K}$, i.e., $34^{\circ}{\rm C}$)
above a gold-coated surface, as illustrated in Fig.~\ref{fig1}a. 
The gold flakes are fabricated using a wet-chemical synthesis method~\cite{chen2016} that produces cetrimonium bromide ligand molecules (CTAB) stabilized single crystalline gold flakes (diameter, $d=3\pm2\,\mu$m) with average thickness of $b=34\pm10\,{\rm nm}$ in aqueous solution \cite{munkhbat2021tunable}
and of various geometrical shapes (SEM image in Fig.~\ref{fig2}a, details in Methods \ref{sec:ExpDet}). 
These flakes are hydrophilic due to the formation of a CTAB bilayer on their surface \cite{Mitamura2007}.
For this study, we select flakes with the shape of equilateral hexagons (inset of Fig.~\ref{fig2}b) due to their larger degree of symmetry.

The substrate is constituted by a glass slide on which a 35-nm-thick gold coating is deposited by sputtering.
In order to control the surface  preferential adsorption, the gold coating is chemically functionalized by a SAM of either hydrophilic or hydrophobic thiols by immersing the surface in a 1-mmol thiol solution with different end groups (Methods~\ref{sec:ExpDet})~\cite{notsu2005}. 
We tune the temperature of the sample via a two-stage feedback temperature controller with $\pm20\,{\rm mK}$ stability \cite{paladugu2016,magazzu2019}and thus control the strength of the critical Casimir forces.

As anticipated by our theoretical predictions (Fig.~\ref{fig1} and discussion above), a change in the magnitude or sign of the critical Casimir force acting on the flake alters its equilibrium height $h$ above the surface. However, these changes are of the order of few tens of nanometers and, thus, are difficult to measure directly.
A more convenient approach consists in measuring the lateral diffusion of the flakes along the substrate, which depends sensitively on their height: the higher (lower) the flake is above the surface, the larger (smaller) its diffusion is.
Accordingly, we record a video of the flake's motion for a long period of time (ca. $100\,{\rm s}$, with a sample rate of $100\,{\rm Hz}$) using a brightfield microscope (Supplementary Fig.~S2). 
Using digital video microscopy, we reconstruct the projection of the flake's position on the $xy$-plane, where $x$ and $y$ are the Cartesian coordinates along the surface of the substrate (solid line in Fig.~\ref{fig2}b, Supplementary Video 1). 
The flake is freely diffusing along the $xy$-plane, while its motion along the  $z$-direction (perpendicular to the substrate) is negligible. 
From the $xy$-trajectory, we calculate the mean square displacement (MSD($t$) at time $t$ in Fig.~\ref{fig2}c) which, for a freely diffusing flake, grows as ${\rm MSD}(t) = 4 D t$ where $D$ is the diffusion constant.
$D$ can be obtained by fitting the measured MSD (Methods~\ref{sec:Diff}). 
Since $D$ depends sensitively on $h$, its value can be used to infer the equilibrium position $h$ of the flake above the surface. 
We have calibrated the theoretical reference $D(h)$ (blue line in Fig.~\ref{fig2}d) using hydrodynamic simulations of the flake (Methods~\ref{sec:sim}), 
which permits a determination of the equilibrium height of the flake with nanometer precision ($\pm1.3\,{\rm nm}$), as shown by the black dashed lines in Fig.~\ref{fig2}d. 
In addition, we validated the indirect measurement of $h$ obtained as described above by alternatively measuring the Fabry-P\'erot cavity modes (Section II of Supplementary Information). 
In fact, a microcavity is formed between the flake and the gold substrate, of which we measure the reflectivity spectrum to determine the spatial variation of the cavity thickness using a spectrometer (Methods \ref{sec:optic} and Supplementary Fig.~S3)\cite{munkhbat2021tunable}. 
The measured cavity modes in the reflection spectrum at different positions were analyzed by the standard transfer-matrix method \cite{munkhbat2021tunable}. We find that the flake is about $90\pm10\,{\rm nm}$ away from the surface inside a critical water--2,6-lutidine mixture at $\Delta T=-0.5\,{\rm K}$, in good agreement with our measurements $h=100\,{\rm nm}$ using hydrodynamic simulations.

\subsection{Measurement of the Casimir--Lifshitz force}

\begin{figure*}[h!]
    \includegraphics[width=1\textwidth]{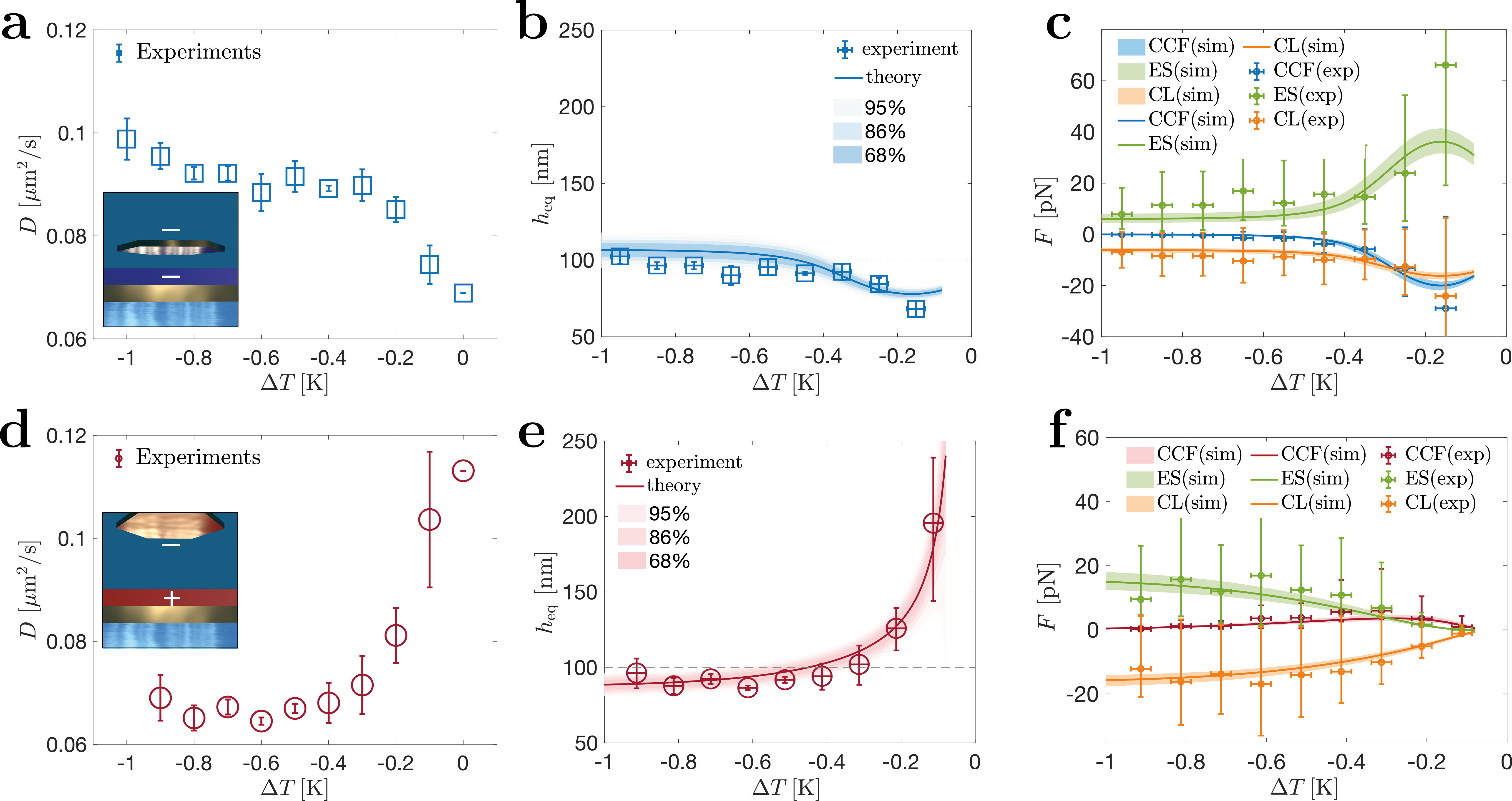}
    \caption{\footnotesize
    {\bf Critical Casimir forces overcome Casimir--Lifshitz force.} 
    {\bf a} The diffusion constant $D$ of a hydrophilic hexagonal flake with $a=700\,{\rm nm}$ above a hydrophilic surface (\emph{symmetric} $(-,-)$ boundary conditions) decreases as the temperature $T$ approaches the critical temperature $T_{\rm c}$ (blue squares), while 
    {\bf b} its height $h$ decreases (blue squares).
    The solid blue line is the best fit of the theoretical model (Eq.~\eqref{eqn:Utot}), from which we extracted $\left\langle h \right\rangle$ in Eq.~\eqref{eqn:hmean}. 
    {\bf c} The flake is under the action of the Casimir--Lifshitz force (orange dots) and the critical Casimir force (blue dots), which are both attractive and whose strengths increase as the flake approaches the surface, while electrostatic forces (green dots) are repulsive. The orange and blue solid lines are the theoretical fitting of critical Casimir (Eq.~\ref{eqn:CCF}) and Casimir--Lifshitz (Eq.~\ref{eqn:CF}) forces, respectively. Note how the critical Casimir force becomes stronger than to the Casimir--Lifshitz force at $\Delta T \approx -0.3\,{\rm K}$.
    {\bf d} The diffusion constant $D$ (red circles) for a hydrophilic hexagonal flake with $a = 840\,{\rm nm}$ above a hydrophobic surface (\emph{antisymmetric} $(-,+)$ boundary conditions) increases upon approaching $T_{\rm c}$, as {\bf e} its height $h$ (red circles) increases.
    The solid red line is the best fit of the theoretical model (Eq.~\eqref{eqn:Utot}) of the average height  $\left\langle h \right\rangle$ (Eq.~\eqref{eqn:hmean}) under the action of {\bf f} the Casimir--Lifshitz force (orange dots, attractive), the critical Casimir force (red dots, repulsive), and the electrostatic force (green dots, repulsive). The red, orange, and green solid lines are the theoretical values from the fit of critical Casimir, Casimir--Lifshitz, and electrostatic forces, respectively.
    The error bars on the experimental points in {\bf b},{\bf e} are the standard deviation from three measurements. The error range for the theoretical lines in {\bf b}, {\bf e} are indicated as shaded areas for the confidence levels of $68\%$, $86\%$, and $95\%$.
    In {\bf c} and {\bf f}  only the $68\%$ confidence level error range are represented.
    }
    \label{fig3}
\end{figure*}

Using the method described in the previous section, we can investigate the separate effects of the various forces on the flake. 
We start by determining the (attractive) Casimir--Lifshitz force at low temperature $\Delta {T} \approx -1\,{\rm K}$, at which the critical Casimir force is negligible (importantly, the Casimir--Lifshitz force and the electrostatic force are not significantly affected by a temperature change up to several K). 
The Casimir--Lifshitz force can then be determined by comparing the total force (which includes electrostatics) acting on the gold flake above the uncoated glass surface (where the Casimir--Lifshitz force is weak) and the force acting when the flake is above a gold-coated surface. 
In fact, we observe that the flake suspended above an uncoated glass surface hovers at $h\approx300\,{\rm nm}$ (Supplementary Fig.~S4), 
while its equilibrium height reduces to $h\approx100\,{\rm nm}$, when the flake is floating above the gold-coated surface. 
This fact is largely independent of the gold functionalization and, therefore, of the adsorption properties of the surface (as can be seen from the values of $h$ at low temperature, $\Delta T = -1\,{\rm K}$,  in Figs.~\ref{fig3}b and \ref{fig3}e). This reduction of the value of $h$ is the result of the presence of an attractive Casimir--Lifshitz forces between the gold flake and the gold surface~\cite{garcia2012}.
Further evidence of the nature of this force is provided by the quality of the fit of the experimental data with the theoretical model (Supplementary Fig.~S5). 
In particular, we compare the inferred heights with the theoretical average height of the flake floating above the (i) gold-coated and (ii) uncoated silica surface. This fit allows us to infer the parameters of the electrostatic interaction in the two cases that are important for the theoretical interpretation of the experimental results provided in the following sections (Results \ref{sec:BC} and \ref{sec:proof}).
The details of the fitting procedure are given in Methods \ref{sec:forces}. 

\subsection{Interplay between critical Casimir and Casimir--Lifshitz forces}
\label{sec:BC}

As we showed in Fig.~\ref{fig1}c, repulsive critical Casimir forces are theoretically expected to be able to overcome Casimir--Lifshitz forces in magnitude as the solvent's temperature approaches $T_{\rm c}$.
In this section, we provide experimental evidence of this fact.

In Figs.~\ref{fig3}a-c, we consider an hexagonal hydrophilic flake with $a=700\,{\rm nm}$ (Fig.~\ref{fig2}b) above a hydrophilic surface, realizing symmetric $(-,-)$ boundary conditions.
As the temperature of the solution is increased towards $T_{\rm c}$, the flake's diffusion constant $D$ decreases from $\approx0.09\,{\rm \mu m^2 s^{-1}}$ to $0.07\,{\rm \mu m^2 s^{-1}}$ (blue squares in Fig.~\ref{fig3}a), indicating that the flake's distance from the substrate decreases from $108\,\pm\, 7\, {\rm nm}$ to $68\,\pm\, 5\,{\rm nm}$ (blue squares in Fig.~\ref{fig3}b). This decrease is primarily due to an increasingly stronger critical Casimir force (blue symbols in Fig.~\ref{fig3}c), but also to an also increase of the Casimir--Lifshitz force (orange crosses in Fig.~\ref{fig3}c) as the equilibrium distance $h$ of the flake decreases.
The experimental results
\footnote{We remark that, in Fig.~\ref{fig3}cf, the experimental points (with error bars) for the various forces have been calculated by inserting the values of $h$ determined experimentally in the expression of the theoretical model for the corresponding forces, given in  
Eqs.~(1-3).
The uncertainties associated with the theoretical predictions of the various forces are calculated considering the interval with $68\%$ confidence level.}
agree well with theoretical predictions for the Casimir--Lifshitz forces (orange solid line in Fig.~\ref{fig3}c), the critical Casimir forces (blue solid line in Fig.~\ref{fig3}c), the electrostatic force (green solid line in Fig.~\ref{fig3}c), as well as the resulting height of the flake (blue solid line in Fig.~\ref{fig3}b) (Methods~\ref{sec:forces}). 
We emphasize that the magnitude of the critical Casimir forces acting on the flake becomes larger than that of the Casimir--Lifshitz forces as the temperature is sufficiently close to $T_{\rm c}$ (i.e., $\Delta T>-0.3\,{\rm K}$) as shown in  Fig.~\ref{fig3}c.

In Figs.~\ref{fig3}d-f, we consider the case of a hexagonal hydrophilic flake with $a=840\,{\rm nm}$ above a hydrophobic surface realizing antisymmetric $(-,+)$ boundary conditions.
Interestingly, while the sizes of the two flakes are different, their equilibrium heights $h\approx100\,{\rm nm}$ far from $T_{\rm c}$ are similar, confirming that the forces acting per unit area are similar and not affected by the critical Casimir contribution, which would have opposite signs for the two flakes.
In this case however, as the temperature approaches $T_{\rm c}$, the flake's diffusion constant $D$ increases (red circles in Fig.~\ref{fig3}d) from $\approx0.067\,{\rm \mu m^2 s^{-1}}$ to $\approx0.15\,{\rm \mu m^2 s^{-1}}$, indicating that its height $h$ above the surface increases (red circles in Fig.~\ref{fig3}e).
In particular $h$ changes from $h\approx 100\,{\rm nm}$ at $\Delta T \approx -1\,{\rm K}$ to $h\approx 200\,{\rm nm}$ at $\Delta T \approx -0.1\,{\rm K}$.
This increase is primarily due to the emergence of repulsive critical Casimir forces (red circles in Fig.~\ref{fig3}f), combined with the decrease of the magnitude of the Casimir--Lifshitz forces (orange circles in Fig.~\ref{fig3}f) as a consequence of the larger distance between the flake and the substrate.
These experimental results fit well with the theoretical predictions (solid lines in Figs.~\ref{fig3}e-f) (Methods~\ref{sec:forces}). 
Also for this case of antisymmetric boundary conditions, the magnitude of the critical Casimir forces eventually exceeds that of the Casimir--Lifshitz force.
Interestingly, at $\Delta T \approx -0.1\,{\rm K}$, the equilibrium height $h$ is increased to a value at which the electrostatic repulsion becomes negligible, the weight of the flake is still much smaller then the Casimir-Lisfhitz attraction  at that distance, and thus the attraction of the Casimir--Lifshitz force is essentially balanced by the repulsive critical Casimir force.
The electrostatic repulsion in the two cases with different boundary conditions is found to be essentially the same by the fitting procedure (Methods~\ref{sec:forces}). 
Thus, this experiment demonstrates that the repulsive critical Casimir forces, as the only tunable repulsive force in the system, can effectively overcome the Casimir--Lifshitz attraction upon approaching the critical temperature $T_{\rm c}$.

\subsection{Spatio-temporal control}
\label{sec:proof}

\begin{figure*}[h!]
    \includegraphics[width=.7\textwidth]{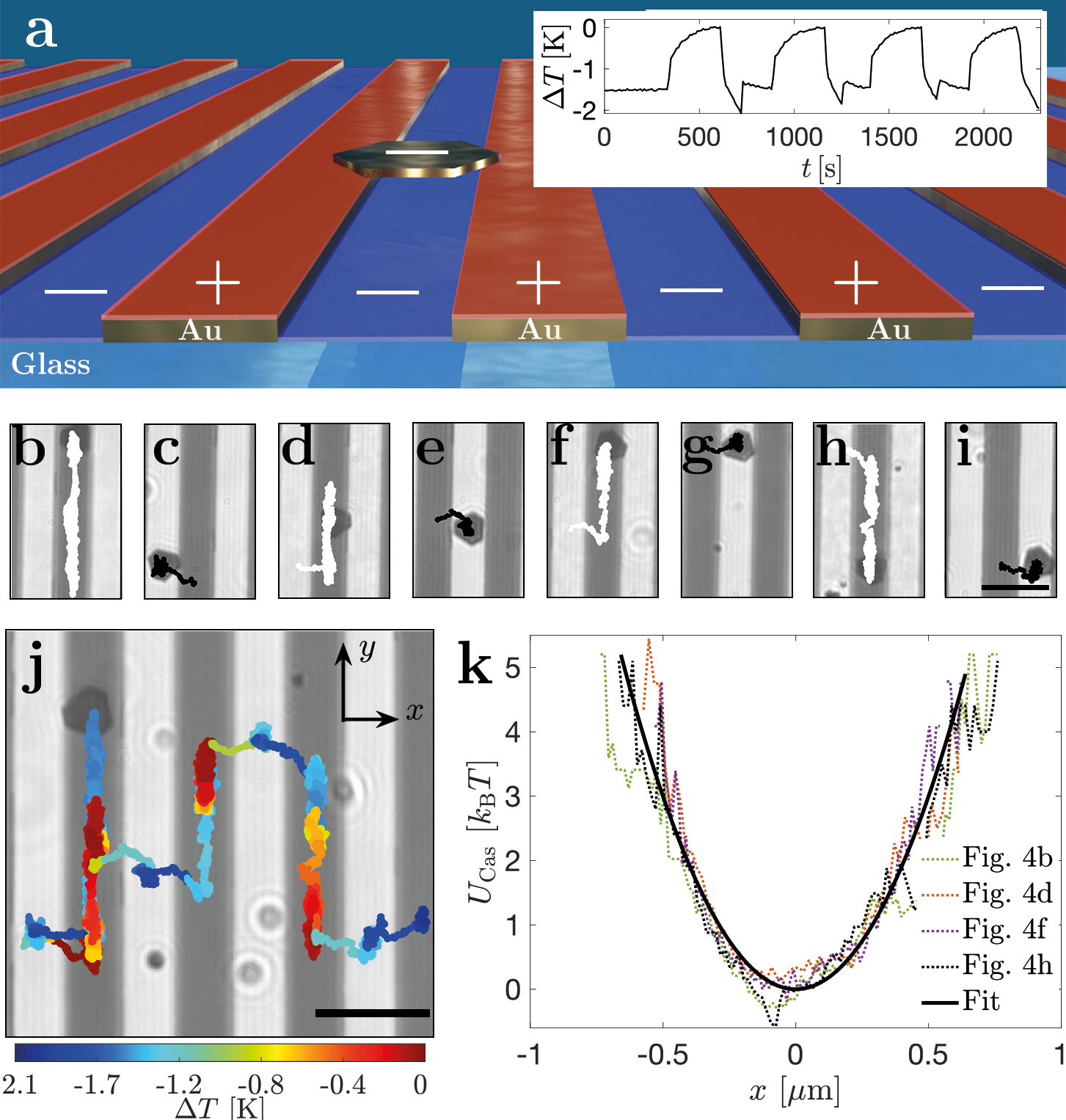}
    \caption{
    {\bf Trapping and release of a flake.} 
    {\bf a} {\em Hydrophobic} gold stripes (thickness $30\,{\rm nm}$, width $3\,{\rm \mu m}$, separation $3\,{\rm \mu m}$) nanofabricated on a {\em hydrophilic} glass substrate.
    Inset: periodic raising and lowering of the temperature $T$ to and away from the critical temperature $T_{\rm c}$.
    {\bf b}-{\bf i} Trajectories of a {\em hydrophilic} gold flake {\bf b},\,{\bf d},\,{\bf f},\,{\bf h} when $T$ is appreciably below $T_{\rm c}$ (white trajectories)
    so that the Casimir--Lifshitz force dominates and confines the flake over the gold stripes (dark stripes in background images), and 
    {\bf c},\,{\bf e},\,{\bf g},\,{\bf i} when $T \approx T_{\rm c}$ (black trajectories) so that the critical Casimir force emerges, repelling the flake from the gold stripes to the glass regions (light stripes in background images).
    \textbf{j} Full trajectory of the flake under periodic heating and cooling of the sample showing that transitions occur shortly after reaching $T \approx T_{\rm c}$ (Supplementary Video V2). 
    Its starting point is marked by the particle image.
    The scale bars represent $6\,{\rm \mu m}$.
    \textbf{k} Casimir--Lifshitz potential along the $x$-direction (see axes in \textbf{j}) in case of a trapped flake above gold stripes. The dotted lines represent the experimentally measured potential for the trajectories in {\bf b}, {\bf d}, {\bf f}, and {\bf h}, while the solid line represents the fitting to a harmonic potential with trap stiffness $k_x\approx 0.1\,{\rm pN}\, \mu{\rm m}^{-1}$.
}
\label{fig4}
\end{figure*}

In the following, we experimentally show how a particle confined to a metallic surface by Casimir--Lifshitz attraction can be lifted away from the substrate by the repulsive critical Casimir force and, therefore, start diffusing freely. For this simple proof-of-principle experiment we fabricated gold stripes (width $3\,{\rm \mu m}$) separated by gaps of glass of the same width, as shown in Fig.~\ref{fig4}a.
We treat the glass stripes and the gold stripes with SAM of opposite wetting properties to make them hydrophilic and hydrophobic, respectively. 
A hydrophilic gold flake is floating above this configuration. The size of the flake is indicated in Fig.~\ref{fig4}a and is comparable to the stripe width.
Accordingly, symmetric boundary conditions $(-,-)$ are realized when the flake is entirely above the glass stripes, while the boundary conditions are antisymmetric $(+,-)$ when it is entirely above the gold stripes. 
In order to study the interplay between the Casimir--Lifshitz forces and the critical Casimir forces, the temperature of the mixture is modulated periodically in time as shown in the inset in Fig.~\ref{fig4}a.
At temperatures far below $T_{\rm c}$, i.e. with $\Delta T \approx - 1.5\,{\rm K}$, the flake is essentially confined on the gold stripes by the Casimir--Lifshitz forces, as shown by its trajectory (white line in Fig.~\ref{fig4}b). 
In fact, one experimentally observes that the flake cannot escape from the gold stripe even at long times, up to tens of minutes. This behavior reveals the presence of an energy barrier that prevents the flake from moving to the neighboring glass stripes.
In agreement with the results shown in section~\ref{sec:BC},
we know that the flake hovers at an equilibrium height $h \approx 100\, {\rm nm}$ over the gold substrate for $\Delta T \lesssim -0.5\,{\rm K}$.
Instead, when $T$ is increased towards $T_{\rm c}$, repulsive critical Casimir forces emerge and lift the flake away from the gold stripe, similar to what we have experimentally observed in Fig.~\ref{fig3}e.  
Thus, the flake freely diffuses away from the gold stripe and, eventually, reaches the glass stripe. Here, due to the symmetric boundary conditions it is attracted towards the glass by the presence of attractive critical Casimir forces (Fig.~\ref{fig4}c).
Only after decreasing $T$ far below $T_{\rm c}$, does the critical Casimir force diminish sufficiently for the flake to get attracted and trapped onto a gold stripe again.

This cycle can be repeated by adjusting the temperature of the system, resulting in continuous transitions between gold and glass stripes, as shown in Figs.~\ref{fig4}b-i and in Supplementary Video~2. 
The complete trajectory is shown in Fig.~\ref{fig4}j, which is color-coded based on the measured temperature:
It can be seen that the transitions between adjacent stripes only occur after $T_{\rm c}$ (dark red) has been reached.

%

This experimental behavior can be explained by a theoretical model explained in details in Section III of the Supplementary Information. 
In the theoretical model describing a flake suspended on a substrate coated with gold stripes, 
the flake is subject to a potential $U(\Delta T; x, h)$ (see Supplementary Fig.~S6a-h), 
which is a periodic function of the coordinate $x$ of the center of the flake along the direction perpendicular to the stripes.
As discussed in details in Methods \ref{sec:forces}, 
the dynamics of the suspended flake along the $x$ direction for a fixed $\Delta T$ is ruled by an effective potential $U_{\rm eff}(\Delta T; x)$ (see Supplementary Fig.~S6i-p). 
For $\Delta T \lesssim-1\, {\rm K}$, the minimum of $U_{\rm eff}$ is located in the middle of the gold stripe (Supplementary Fig.~S6i), 
where the probability density distribution $P_{\rm eff}(\Delta T; x)$ has a sharp peak (Supplementary Fig.~S6q). 
Upon decreasing $\Delta T$ the peak of $P_{\rm eff}$ becomes less pronounced, though remaining localized at the center of the gold stripe. Eventually, for $\Delta T\approx -0.1\, {\rm K}$, the probability distribution develops two comparable peaks (Supplementary Fig.~S6v), the second being localized at the center of the silica stripe. For $\Delta T \gtrsim -0.08\, {\rm K}$, this second peak becomes predominant (Supplementary Fig.~S6x), in agreement with previous results in the literature \cite{soyka2008,trondle2009normal,trondle2011molecular}.

The stiffness of the lateral Casimir--Lifshitz force trapping the flake on the gold stripe when $T \ll T_{\rm c}$ can be quantified by measuring the effective attractive potential to which the flake is subject.
This is done by calculating the probability distribution $P_{\rm exp}(x)$ of the $x$-component of the trajectories shown in Figs.~\ref{fig4}b,\,d,\,f,\,h from which the effective potential $U_{\rm exp}(x)$ is determined by inverting the Boltzmann factor, i.e., $U_{\rm exp}(x) = -k_{\rm B} T \ln{\left( P_{\rm exp}(x)\right)}$.
The resulting experimental potentials are shown by the dashed lines in Fig.~\ref{fig4}k, which are then fitted to a harmonic potential $U(x) = {k\over2}x^2$ (black solid line in Fig.~\ref{fig4}k) with stiffness $k \approx 0.1\,{{\rm pN}\,\mu {\rm m}^{-1}}$.
Comparing with the theoretical model (Supplementary Fig.~S7), 
we see that we can obtain a value of the stiffness of the same order of magnitude for an appropriate range of the flake's size. 

\section{Conclusions}
\label{sec:Concl}

We provided here a direct experimental evidence of the fact that critical Casimir forces can be employed to counteract the attraction due to Casimir--Lifshitz forces.
We demonstrated this by studying the average height in equilibrium of a gold flake above different surfaces (gold, glass) with various preferential adsorptions (hydrophilic, hydrophobic) for the two components of the water--2,6-lutidine mixture.
We determined the height of the flake above the surface, with tens of nanometer precision, by measuring its diffusion constant.
In a simple proof-of-principle configuration, we have demonstrated that we can control the trapping of a metallic flake due to the lateral Casimir--Lifshitz attraction and reversibly switch between trapping and release by tuning the temperature of the critical binary liquid mixture where the flake is dispersed, achieving a precise spatio-temporal control of its position in space.
Our method provides a novel way of controlling the distances of metallic micro- and nanostructures using tunable critical Casimir forces to counteract forces such as the Casimir--Lifshitz force, thereby preventing stiction and device failure.
Moreover, this path opens up new possibilities for the dynamic control of MEMS and NEMS where the temperature of the system could be controlled via light illumination, enabling faster transitions and higher selectivity for a new generation of micromembranes ubiquitously found in MEMS and NEMS devices. 

\section{Methods}
\label{sec:meth}

\subsection{Calculation of the Casimir--Lifshitz and critical Casimir forces}
\label{sec:forces}

The theoretical model which relates the measured experimental temperatures to the  equilibrium heights of the flake in the presence of two different boundary conditions of the order parameter, accounts for the following forces: weight and buoyancy of the flake,
DLVO electrostatic interaction, Casimir--Lifshitz force, and critical Casimir force.

{\bf Weight and buoyancy.} For the hexagonal flakes used in our experiments (Fig.~\ref{fig2}a) the sum of the weight and the buoyancy is within the range of  $5$ to $30\, {\rm fN}$,
depending on the volume of the flake. 
In the case of a flake suspended above a gold-coated substrate, the sum of weight and buoyancy is negligible in comparison with the Casimir--Lifshitz attraction and the electrostatic forces, the magnitude of which is within the range $\approx 1$ to $10\, {\rm pN}$ for distances of the order of $\approx 100$ to $200\, {\rm nm}$, such as those observed in our experiments. 
In the case of a flake suspended above an uncoated silica substrate, instead, the sum of weight and buoyancy is not negligible compared to the other relevant forces because the flake hovers at a higher $h$ and the Casimir--Lifshitz attraction is $\approx 20$ times smaller (for the same $h$).
Hence, we have included weight and buoyancy in all theoretical modelling, even though omitting their contribution for the gold-coated substrate does not change the results for the explored experimental parameters.

{\bf Electrostatic force.} The DLVO electrostatic force\cite{israelachvili2011intermolecular} $F_{\rm ES}$ between the flake and the substrate, at a distance $h$, is modelled as
\begin{equation}\label{eqn:ESforce}
	\frac{F_{\rm ES}(h)}{S} = P_0\,e^{-h/\lambda_{\rm D}},
\end{equation}
where $S$ is the area of the hexagonal flake, $\lambda_{\rm D}$ is the Debye screening length of the critical mixture, and $P_0$ is a parameter with the dimension of a pressure. 
For each experimental condition we fitted the parameters independently, choosing the combinations of $\lambda_{\rm D}$ and $P_0$ that minimises the $\chi^2$ when comparing the experimental data for $\left\langle h \right\rangle$ with the model. 
We have found the following combinations:
\begin{equation}\label{eqn:listESparameters}
    \begin{array}{| l | l | l |} \hline
       \mbox{gold-coated $(-,-)$}  & \lambda_{\rm D} = 16 \pm 0.5\, {\rm nm} & P_0 = 3.7\pm 0.6\, {\rm kPa}\\ \hline
       \mbox{gold-coated $(+,-)$}  & \lambda_{\rm D} = 17 \pm 0.5\, {\rm nm}& P_0 = 1.5 \pm 0.3\, {\rm kPa} \\ \hline
       \mbox{uncoated silica $(-,-)$}  & \lambda_{\rm D} = 17 \pm 0.5 \, {\rm nm} & P_0 = 28 \pm 5\, {\rm kPa} \\ \hline
    \end{array}
\end{equation}
As expected, these three independent fits give very similar values of $\lambda_{\rm D}$, while the value of $P_0$ in the case of uncoated silica is significantly larger than in the case of gold-coated substrate, indicating a stronger electrostatic repulsion. 

{\bf Casimir--Lifshitz force.} The Casimir--Lifshitz force $F_{\rm CL}$ in Eq.~\ref{eqn:CF} has been calculated according to Ref.~\citenum{parsegian2006} for the layered planar bodies involved in the experimental setup in Fig.~\ref{fig1}a. The corresponding free energy per unit area is given by
\begin{equation}\label{eq:freenergy_2sl_short_goldcoated}
    \begin{array}{rcl}
        \displaystyle G_{\rm CL}(h) & = & \displaystyle -\frac{A_{{2}/{3}}(h)}{12 \pi {h}^2} 
        \displaystyle -\frac{A_{{1}/{3}}(h+a_1)}{12 \pi {(h+a_1)}^2} 
        \displaystyle -\frac{A_{{2}/{4}}(h+d)}{12 \pi {(h+d)}^2} 
        \displaystyle -\frac{A_{{1}/{4}}(h+a_1+d)}{12 \pi {(h+a_1+d)}^2}, \\[14pt]
    \end{array}
\end{equation}
where $A_{{2}/{3}}$, $A_{{1}/{3}}$, $A_{{2}/{4}}$, and $A_{{1}/{4}}$ are the Hamaker's functions, $a_1$ is the thickness of the gold layer on the substrate, and $d$ is the thickness of the gold flake.
In the simpler case of a bottom slide of uncoated silica, instead, the dispersion forces can be derived from
\begin{equation}\label{eqn:freenergy_2sl_short_uncoated}
    \begin{array}{rcl}
        \displaystyle G_{\rm CL}(h) & = & \displaystyle -\frac{A_{{5}/{3}}(h)}{12 \pi {h}^2} 
        \displaystyle -\frac{A_{{5}/{4}}(h+d)}{12 \pi {(h+d)}^2}, \\[14pt]
    \end{array}
\end{equation}
where $A_{{5}/{3}}$ and $A_{{5}/{4}}$ are the corresponding Hamaker's functions.
See also Section I of Supplementary Information. 

{\bf Critical Casimir force.} 
The expression of the critical Casimir force $F_{\rm crit}$ in Eq.~\eqref{eqn:CCF} depends on the correlation length $\xi$ of the order parameter fluctuation of the binary liquid mixture, which is related to $\Delta T$ as $\xi = \xi_0 \left( \frac{\Delta T}{T_{\rm c}} \right)^{-\nu}$, where $\xi_0 \approx 0.22\,{\rm nm}$ for the water--2,6-lutidine mixture while $\nu \backsimeq 0.63$ is the  critical exponent of the Ising universality class to which the mixture belongs \cite{gambassi2009critical}.

{\bf Total force and potential.} The total force $F_{\rm tot}$ acting on a flake is
\begin{equation} \label{eqn:Ftot}
    F_{\rm tot}(h,\Delta T) = F_{\rm ES}(h) + F_{\rm CL}(h) + F_{\rm crit}(h,\Delta T) + F_{\rm w+b},
\end{equation}
where the term $F_{\rm w+b}$ due to weight and buoyancy can be neglected in the case of a flake suspended on a gold-coated substrate at the typical distances observed in our experiments, as explained above.
The total potential $U_{\rm tot}$ of the flake is therefore:
\begin{equation} \label{eqn:Utot}
    U_{\rm tot}(h,\Delta T) = U_{\rm ES}(h) + G_{\rm CL}(h) \cdot S + U_{\rm crit}(h,\Delta T) + U_{\rm w+b}.
\end{equation}
Such potential is necessarily asymmetric, being defined only for $h>0$. In the case of a gold flake floating on a gold-coated substrate, $U_{\rm tot}(h,\Delta T)$ displays a very sharp minimum at a certain $h(\Delta T)$ for all $\Delta T$,  while for the flake suspended on an uncoated silica substrate the minimum of the potential becomes shallow for $\left| \Delta T \right| \gtrsim 0.1\, {\rm K}$. 

In comparing with the experiments, one has to keep in mind that the experimentally accessible quantity is the average height of the potential $\left\langle h \right\rangle$.

\begin{equation}\label{eqn:hmean}
    \left\langle h \right\rangle = \int_{0}^{+\infty}  {\rm d}h\, h\, P(h,\Delta T),
\end{equation}
which is defined in terms of the equilibrium probability distribution function $P(h,\Delta T)$ associated with the potential $U(h,\Delta T)$:
\begin{equation}
    P(h,\Delta T) = \frac{1}{Z} e^{-\frac{U(h,\Delta T)}{k_{\rm B} T}},
\end{equation}
where $Z = \int_{0}^{+\infty} e^{-\frac{U(h,\Delta T)}{k_{\rm B} T}}\, {\rm d} h $. 
Hence, the comparison with the experimental data must consider the equilibrium height of the potential $h$, defined as:

\begin{equation}\label{eqn:hequi}
\frac{{\rm d}U (h_{\rm eq})}{{\rm d}h} = 0.
\end{equation}
In a potential with a sharp dip, $h$ and $\left\langle h \right\rangle$ are often very close, and therefore $\left\langle h \right\rangle$ can be estimated by $h$.
In contrast, in a shallow, asymmetric potential, the equilibrium height $h$ can differ significantly from the average height $\left\langle h \right\rangle$.
In the experimental conditions considered here, the difference between $\left\langle h \right\rangle$ and $h$ for a gold flake suspended on an uncoated silica surface is $| \left\langle h \right\rangle - h_{\rm eq} | \approx 50-100 {\rm nm}$, while when the flake is suspended on a gold substrate this difference is $< 5\, {\rm nm}$.

In order to compare the predictions of the model discussed above with the experimental data, we focus on the average height $\left\langle h \right\rangle$ due to the total potential considering also the amplitude $\Delta h$ of its fluctuations. In our calculation and fitting procedure, $\Delta h$ is set to correspond to a confidence level of $68\%$. 
Only $F_{\rm crit}(h)$ and $U_{\rm crit}(h)$ depend significantly on the minute variations of $\Delta T$ (via the dependence on $\xi(t)$ of the scaling function $\theta\left({h}/{\xi} \right)$, Eq.~\eqref{eqn:CCF}). The electrostatic and Casimir--Lifshitz interactions do not depend on $\Delta T$, with the caveat that, when the correlation length $\xi$ of the solvent becomes comparable with the Debye screening length $\lambda_{\rm D}$ of the electrostatics, other effects might come into play. 
In fact, when $\lambda_{\rm D}$ becomes comparable with $\xi$, the description of the electrostatics with Eq.~\eqref{eqn:ESforce} is not appropriate anymore. 
Given that the estimated value of the Debye screening length is $\lambda_{\rm D} \backsimeq 16-17\, {\rm nm}$, the value of the corresponding correlation length $\xi$\footnote{For $\Delta T \lesssim 0.1\, {\rm K}$ for the correlation length we have $\xi \lesssim 10 \mathrm{nm}$, which is smaller than the value we have for $\lambda_{\rm D}$, i.e., $16$--$17\, \mathrm{nm}$ } is smaller than $\lambda_{\rm D}$ for the typical temperature differences in our system up to $\Delta T \lesssim 0.1\, {\rm K}$.

We first determined the parameters of the electrostatic interaction by fitting the value of the Debye screening length $\lambda_{\rm D}$ and the prefactor $P_0$ such that they minimize
\begin{equation}\label{eqn:chi2}
    \chi^{2} = \sum_{j=1}^{N} \frac{\left( \left\langle h_{j,\rm exp} \right\rangle - \left\langle h_{j} \right\rangle \right)^{2}}{\Delta h_{j}^{2}},
\end{equation}
where $\left\langle h_{j} \right\rangle$ and $\Delta h_{j}$ were calculated for each different combination of parameters for the experimental temperature $\Delta T_{j}$ of the experimental data $\left\langle h_{j,\rm exp} \right\rangle$.
The fitted $\lambda_{\rm D}$ is in the range of values obtained in similar experiments \cite{paladugu2016,magazzu2019}. 

\subsection{Experimental details}
\label{sec:ExpDet}

We consider a dilute suspension of gold flakes (monocrystalline, polydisperse, and with average thickness $ d = 34\pm10$\,nm determined via AFM measurements \cite{munkhbat2021tunable}). The gold nanoflakes are wet-chemically synthesized using a rapid and seedless method in aqueous solution described in details in Ref.~\citenum{chen2016}. Briefly, $100\,{\rm \mu }$l of $100\,{\rm mM\, HAuCl_4}$ is added to $3\,{\rm ml}$ of $20\,{\rm mM\,CTAB}$ aqueous solution in a glass vial, and the mixture is gently mixed and left undisturbed for several minutes. Then, $100\,{\rm \mu l}$ of $100\,{\rm mM}$ ascorbic acid is added to the mixture, followed by rapid inversions of the vial for $10\,{\rm s}$. The resulting solution is immediately placed in a water bath at $85^{\circ}{\rm C}$ and kept undisturbed for about an hour. The products are washed by centrifugation at $4000\,{\rm rpm}$ for $10\,{\rm min}$ and finally dispersed in deionized water for further experiments.

The flakes obtained as described above are then suspended in a near-critical binary liquid mixture of water and 2,6-lutidine at the critical composition of lutidine $c^{\textrm{c}}_{\textrm{L}}=0.286$, having a lower critical point at $T_{\textrm{c}}\approx34^{\circ}$C \cite{Grattoni1993} (Supplementary Fig.~8). 
The suspension is confined in a sample cell between a microscopy slide and a cover slip spaced by about $300\,{\rm \mu m}$.

On top of the cover slip, a 40-nm-thick gold layer is sputtered homogeneously across the sample. The cover slip is then left overnight in a 1-mM solution of thiols (1-Octanethiol for hydrophobic and 6-Mercapto-1-hexanol for hydrophilic treatment) and ethanol, creating a self-assembled monolayer (SAM) on top of the gold.
To create a hydrophobic SAM on top of the glass, Trichloro(1H,\ 1H,\ 2H,\ 2H-perfluorooctyl)silane (PFOCTS) has been evaporated for 4\,h under vacuum. The stark contrast in the resulting wetting angle between hydrophilic SAM ($\theta=19^{\circ}$, measured from the side) and hydrophobic SAM ($\theta=102^{\circ}$) of a 10-${\rm \mu l}$ water droplet can be seen in Supplementary Fig.~S9. 

For the patterned substrate shown in Fig.~\ref{fig4}, an additional 3-nm-thick layer of titanium has been added to the glass substrate for better adhesion before sputtering gold. 
The patterned gold stripes are fabricated by direct laser-writing and lift-off. 
The substrates are spin coated with LOR3A ($4000\,{\rm rpm}$ for $60\,{\rm s}$, baking at $200^{\circ}{\rm C}$ for $5\,{\rm min}$) and S1805 ($3000\,{\rm rpm}$ for $45\,{\rm s}$, baking at $110^{\circ}{\rm C}$ for $1\,{\rm min}$). The samples are exposed using a Heidelberg DWL2000 direct laser writer and developed in MF CD26 for $50\,{\rm s}$. Titanium and gold are deposited and lift-off is performed in hot remover 1165 at $70^{\circ}{\rm C}$ overnight. 

A schematic of the experimental setup is shown in Supplementary Fig.~S2. 
The whole sample is temperature-stabilized by using a two-stage temperature controller consisting of a copper-plate heat exchanger coupled to a circulating water bath at $T_0=32.5^{\circ}{\rm C}$ (T100, Grant Instruments) and two Peltier elements (TEC3-6, Thorlabs) attached to the imaging objective and in feedback with a controller unit (TED4015, Thorlabs) reaching $\pm20\,{\rm mK}$ temperature stability \cite{paladugu2016,magazzu2019}.
The flake's translational motion is captured using digital video microscopy at a frame rate of $100\,{\rm Hz}$. The particle images are thresholded to determine the particle's centroid. The reconstructed trajectory is then analyzed using standard methods\cite{crocker1996methods,jones2015optical}.

\subsection{Diffusion of a flake}
\label{sec:Diff}

The determination of the equilibrium height $h$ of a flake depends on three crucial factors: the measurement of the diffusion constant $D$, the hydrodynamic simulations for extracting $h$ from $D$, and the presence of spurious concentrations of salt in solution, as discussed below.
In principle, the diffusion constant $D$ depends on the flake diffusion in three spatial dimensions.
However, as the flake has a mass density $\rho_{\rm Au} \approx 17\cdot 10^{3}\, {\rm kg}\,{\rm m^{-3}}$ much larger than that of the surrounding fluid ($\rho_{\rm WL} \approx 0.98\cdot 10^{3}\, {\rm kg}\,{\rm m^{-3}}$),   
it settles quickly from bulk into proximity of the substrate at an equilibrium distance $h$
so that, in practice, we observe its motion along the horizontal $xy$-plane only, whereas any motion along the vertical direction $z$ is negligible: a theoretical estimate based on the potential we obtain with our model (Methods~\ref{sec:forces}) 
gives that a typical amplitude of the the longitudinal fluctuation $\Delta h$ is within $\Delta h \lesssim 5\, {\rm nm}$. 

The presence of salts in solution, which further screen the flake, might reduce the repulsive electrostatic forces and therefore its heights above surface, as it was previously studied \cite{munkhbat2021tunable}.
However, we can neglect any influence of salt in our system, as our gold flakes, originally prepared in CTAB buffer solution, have been diluted more than 2000 times with deionized water. This is confirmed by the fact that, even after the surface treatment with SAM or in control experiments in pure water, the average height $h\approx100\,{\rm nm}$ of the flake does not change (Supplementary Fig.~S4). 

\subsection{Hydrodynamic Simulations}
\label{sec:sim}

In our hydrodynamics simulations, a flake is modelled as a rigid assembly of $N$~spheres (typically, $N\approx 1000$ to $3000$) of radius $r$, with $r=b/2$ where $b$ is the thickness of the flake, glued together and arranged on a regular hexagon. 
The assembled object is immersed in an incompressible Newtonian fluid bounded by a no-slip wall. 
At low Reynolds numbers, the fluid dynamics is governed by the Stokes equations in which viscous forces dominate inertial forces~\cite{happel12}. 
Under these conditions, the translational velocity $\vect{V}_i$ of the sphere labelled by the index $i=1,\dots ,N$ and composing the conglomerate is linearly related to the force $\vect{F}_j$ exerted on either the same sphere  ($j=i$) or on adjacent spheres via the hydrodynamic mobility tensor~\cite{kim13}.
Specifically,
\begin{equation}
	\vect{V}_i = \boldsymbol{\mu}^\mathrm{S} \boldsymbol{\cdot} \vect{F}_i
	           + \sum_{j \ne i} \boldsymbol{\mu}^\mathrm{P} \boldsymbol{\cdot} \vect{F}_j \, , 
\end{equation}
where $\boldsymbol{\mu}^\mathrm{S}$ denotes the self-mobility tensor of a single sphere sedimenting near a hard wall and $\boldsymbol{\mu}^\mathrm{P}$ corresponds to the pair-mobility tensor quantifying the hydrodynamic interactions between the spheres and the wall~\cite{swan07}.
Here, we employ the well-established Rotne-Prager approximation~\cite{rotne69} combined with the Blake solution~\cite{blake71} to account for the corrections due to the presence of the wall.
The diffusion coefficient is obtained from the drag coefficient, which, in turn, is obtained by following the usual procedure of evaluating the mean drag per sphere within the assembly~\cite{tseng2001translation}.
The accuracy of our approach has been assessed by close comparison of the predicted bulk drag coefficients with the exact numerical simulations of a closely packed conglomerate of spheres which was obtained by using the freely available and open-source library HYDROLIB~\cite{hinsen1995hydrolib} (Supplementary Fig.~S10). 

These results allow us to simulate the diffusion of a regular hexagonal flake with side length $a$ at height $h$ above a wall as in the representative case of a flake with $a=840\,$nm in Fig.~\ref{fig2}d (cases of particles with different sizes are shown in Supplementary Fig.~S11). 

Moreover, hydrodynamic simulations allow us to investigate the influence of a possible Brownian rotation on the diffusion of the flake in the presence of the confining interface, which we find to be negligible.
As a last remark about the hydrodynamic simulation, its underlying assumption is that the viscosity of the solvent does not depend significantly on the temperature. This is actually the case in experiments: in fact, the viscosity of water-lutidine at critical concentration does not change appreciably in the temperature range of our experiments\cite{stein1972tracer,clunie1999interdiffusion}. 
Moreover, we do not possess experimental evidence of the presence of a wetting layer.

\subsection{Optical measurements and analysis}
\label{sec:optic}
Reflection spectra at normal incidence ($\rm NA=0.5$) were collected by using an inverted microscope (Nikon Eclipse TE2000-E)  equipped with an oil-immersion 100$\times$ objective (switchable $\rm NA=0.5$ to $1.3$, Nikon), directed to a fiber-coupled spectrometer (Andor Shamrock SR-303i), equipped with a CCD detector Andor iDus 420 (see schematic in Ref.~\onlinecite{munkhbat2021tunable}). Obtained reflection spectra were analyzed by the standard transfer-matrix method, described further in Ref.~\onlinecite{munkhbat2021tunable}.

\begin{acknowledgments}
This work was partially supported by the ERC Starting Grant ComplexSwimmers (grant number 677511) and by Vetenskapsr{\aa}det (grant number 2016-03523). AG acknowledges support from MIUR PRIN project ``Coarse-grained description for non-equilibrium systems and transport phenomena (CO-NEST)'' n. 201798CZL. Fabrication in this work was partially done at Myfab Chalmers.
ADMI gratefully acknowledges support from the DFG (Deutsche Forschungsgemeinschaft) through the project No. DA 2107/1–1.
\end{acknowledgments}

\section{Author contributions}
All authors have discussed the results and have written or edited the manuscript. FS, BM and GV planned the experiments. FS and BM performed the experiments. ADMI and HL conducted the hydrodynamic simulations. AC and AG conceptualized the theoretical model and performed the force simulations. FS and RV fabricated the experimental samples. BM fabricated the flakes.

\section{Additional Information}
\vskip 0.3cm 
\noindent\textbf{Competing interests:}
\\The authors declare no competing interests.
\vskip 0.3cm 
\noindent\textbf{Data availability:}
\\All data are available from the corresponding authors upon reasonable request.
\vskip 0.3cm 
\noindent\textbf{Code availability:}
\\The codes that support the findings of this study are available from the corresponding authors upon reasonable request.
\vskip 0.3cm 
\noindent\textbf{Correspondence:}
\\Correspondance and requests for materials should be addressed to FS or GV. 


\bibliographystyle{naturemag}
\bibliography{CL_CCF_arxiv}

\end{document}